\documentclass[a4paper]{jpconf}
\usepackage{graphicx}
\usepackage{tikz}
\usepackage{units}
\usetikzlibrary{decorations.pathreplacing}

\begin{document}
\title{A hybrid deep learning approach to vertexing}

\author{
   Rui Fang\textsuperscript{1},
   Henry F Schreiner\textsuperscript{1,2},
   Michael D Sokoloff\textsuperscript{1},
   Constantin Weisser\textsuperscript{3} and
   Mike Williams\textsuperscript{3}
}

\address{
    \textsuperscript{1} University of Cincinnati, Cincinnati, OH, United States
}
\address{
    \textsuperscript{2} Princeton University, Princeton, NJ, United States
}
\address{
    \textsuperscript{3} Massachusetts Institute of Technology, Cambridge, MA, United States
}

\ead{hschrein@cern.ch}

\begin{abstract}
In the transition to Run 3 in 2021, LHCb will undergo a major luminosity upgrade, going from 1.1 to 5.6 expected visible Primary Vertices (PVs) per event, and will adopt a purely software trigger. This has fueled increased interest in alternative highly-parallel and GPU friendly algorithms for tracking and reconstruction. We will present a novel prototype algorithm for vertexing in the LHCb upgrade conditions.
We use a custom kernel to transform the sparse 3D space of hits and tracks into a dense 1D dataset, and then apply Deep Learning techniques to find PV locations. By training networks on our kernels using several Convolutional Neural Network layers, we have achieved better than 90\% efficiency with no more than 0.2 False Positives (FPs) per event. Beyond its physics performance, this algorithm also provides a rich collection of possibilities for visualization and study of 1D convolutional networks. We will discuss the design, performance, and future potential areas of improvement and study, such as possible ways to recover the full 3D vertex information.
\end{abstract}

\section{Introduction}

% TODO: Write this up

The LHCb detector is facing a major upgrade in luminosity in Run 3 in 2021. In each event, the Poisson average expected for the visible Primary Vertices (PVs) will go from 1.1 to 5.6. In addition, the hardware level-0 trigger is being removed in favor of a purely software trigger running at 30 MHz \cite{CERN-LHCC-2014-016}. Work is ongoing to find new algorithms to support the upgrade era software for LHCb.

\begin{figure}
	\centering
	\begin{tikzpicture}[
    scale=1,
    belowbox/.style={draw},
    topbox/.style={draw, fill=white},
    texttop/.style={below},
    track/.style={thick},
    minibox/.style={draw, thick, minimum width=.1cm, minimum height=.1cm},
    miniarr/.style={->, thick},
    conn/.style={-latex, ultra thick}
    ]
    	\begin{scope}
    	    \path [belowbox] (-1.5,-1) rectangle (1.5,.45);
    		\draw [track] (0,-.275) -- (-1.5,-.9);
    		\draw [track] (0,-.275) -- (-1.5,-.6);
    		\draw [track] (0,-.275) -- (-1.5,.2);
    		\draw [track] (0,-.275) -- (1.5,-.3);
    		\draw [track] (0,-.275) -- (1.5,-.6);
    		\draw [track] (0,-.275) -- (1.5,.85);
    		\draw [track] (0,-.275) -- (-.5,-1);
    		\draw [track] (.4, -.6) -- (1.3,-1);
    		\draw [track] (.4, -.6) -- (.7,-1);
    		\path [topbox]  (-1.5,.45) rectangle (1.5,1);
    	    \node at (0,1) [texttop] {Tracking};
            \path [fill=black!30] (0, -.275) circle (.05);
            \path [fill=black!70] (.4, -.6) circle (.05);
    	\end{scope}

    	\begin{scope}[xshift=4cm]
    	    \path [belowbox] (-1.5,-1) rectangle (1.5,.45);
    		\path [fill=black] plot [smooth] coordinates {
    			(-1.5,-1)
    			(-1.1, -.98)
    			(-.9, -.9)
    			(-.5, -.6)
    			(0, .2)
    			(.2, -.2)
    			(.4, 0)
    			(.8, -.7)
    			(1.1, -.9)
    			(1.3, -.98)
    			(1.5,-1)
    		};
    		\path [topbox]  (-1.5,.45) rectangle (1.5,1);
    	    \node at (0,1) [texttop] {Kernel generation};
    	\end{scope}

    	\begin{scope}[xshift=8cm]
    	    \path [belowbox] (-1.5,-1) rectangle (1.5,.45);

            \begin{scope}[yshift=3mm]
        		\node [minibox] (mA) at (-1,-.3) {};
        		\node [minibox] (mB) at (-.5,-.3) {};
        		\node [minibox] (mC) at (0,-.3) {};
        		\node [minibox] (mD) at (.5,-.3) {};
        		\node [minibox] (mE) at (1,-.3) {};
        		\draw [miniarr] (mA) -- (mB);
        		\draw [miniarr] (mB) -- (mC);
        		\draw [miniarr] (mC) -- (mD);
        		\draw [miniarr] (mD) -- (mE);
            \end{scope}

    		\path [topbox]  (-1.5,.45) rectangle (1.5,1);
    	    \node at (0,1) [texttop] {Make predictions};
            \node at (0,-.5) {CNNs};
    	\end{scope}

    	\begin{scope}[xshift=12cm]
    	    \path [belowbox] (-1.5,-1) rectangle (1.5,.45);
    		\path [fill=black] plot [smooth] coordinates {
    		(-.15,-1)
    		(0, 0)
    		(.15,-1)
    		};
    		\path [fill=black] plot [smooth] coordinates {
    		(.25,-1)
    		(.4, 0)
    		(.55,-1)
    		};
    		\path [topbox]  (-1.5,.45) rectangle (1.5,1);
    	    \node at (0,1) [texttop] {Interpret results};
    	\end{scope}

    	\begin{scope}[xshift=0cm, yshift=-2.5cm]
    	    \path [belowbox] (-1.5,-1) rectangle (1.5,.45);
    		\path [fill=black!30] plot [smooth] coordinates {
    		(-.15,-1)
    		(0, 0)
    		(.15,-1)
    		};
    		\path [fill=black!70] plot [smooth] coordinates {
    		(.25,-1)
    		(.4, 0)
    		(.55,-1)
    		};
    		\path [topbox]  (-1.5,.45) rectangle (1.5,1);
    	    \node at (0,1) [texttop] {Truth};
    	\end{scope}

    	\draw [conn] (1.5,0) -- (2.5,0);
    	\draw [conn] (5.5,0) -- (6.5,0);
    	\draw [conn] (9.5,0) -- (10.5,0);
    	\draw [conn] (1.5, -2.5) -- (7, -1.1) node [above=-2.5, midway, rotate=13] {Training};
    	\draw [conn] (1.5, -3) -- (11, -1.1) node [above=-1, midway, rotate=11] {Validation};
\end{tikzpicture}
	\caption{The procedure followed in work to convert hits into primary vertices.}
	\label{fig:approach}
\end{figure}
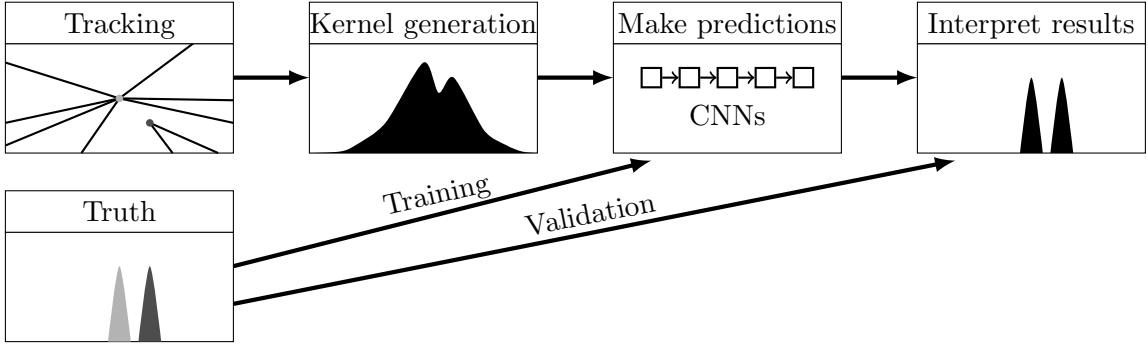

A new method is proposed here for finding vertices from tracks or hits using machine learning techniques (see Figure \ref{fig:approach}). The algorithm starts with tracks containing location, direction, and covariance matrix information. The tracks in 3D space are converted to a 1D binned ``kernel'' through a process described below. Peaks in this kernel are closely related to the z positions of the primary vertices. We use a series of 1D convolutional layers to predict the PV locations from this kernel. The output probabilities from the neural network are converted to a list of PV candidate z-locations using a simple peak finding algorithm. Using this procedure we have surpassed 90\% efficiency with less than 0.2 False Positives (FPs) per event.

A toy simulation of the LHCb detector has been developed to test this algorithm. The design of the algorithm and the toy that it was tested with follows.

\section{Kernel Generation}

Our toy simulation generates tracks from PVs and SVs, and we use a propagation and intersection procedure to produce hits in the Vertex Locater (VELO) according to a simple model of the upgrade detector \cite{Collaboration:1624070}. Pythia \cite{Sjostrand:2006za, Sjostrand:2007gs} creates primary and secondary vertices in the beam interaction region. The particles it produces are then propagated through the detector, with interactions being recorded as hits in the 26 detection planes in the lower and upper halves of the VELO. Scattering in the detector and the foil shielding the VELO from the RF interference produced by the pulsed beam \cite{Collaboration:1624070} are also implemented.

The algorithm presented here applies to tracks, but our simulation produces raw hits, so a simple and somewhat inaccurate procedure is applied; this should be sufficient, since our algorithm is resilient to small tracking errors. We call this system ``prototracking'', and it was originally planned to be primary driver for the kernel generation to provide PV candidates as early as possible in the tracking procedure. Due to recent advances in the tracking code, the full tracks will be available early in the trigger system, so the prototracking system will not be needed in the future. For the prototracking, the hits are radially sorted, and then formed into triplets, with a $\chi^2 < 10$ test for validity. Once a triplet is found, all hits that can be added to the track with $\chi^2<9$ are removed from further searches. This list is built and sent to the kernel.

To provide a simple, dense representation for the machine learning algorithm to use, the tracks are transformed using a ``kernel'' procedure. For any point in space, the value of the kernel can be computed as 
\begin{equation}
\rho \equiv  \frac{ \sum  p^2}{\sum  p} - \sum p,
\end{equation}
where $p$ is Gaussian function in terms of distance to the impact parameter (IP) in x and y of a track, and is summed over all tracks.
In the future, a full covariant matrix will be used, but for now, a constant value for the Gaussian width of 0.05 is used, unless $\chi^2>6$, in which case $ (\chi^2-2)\frac{0.05}{4} $ is added to the width.
The final Gaussian is just the product of the Gaussians in $ \mathrm{IP} \cdot \hat{x}$ and $ \mathrm{IP} \cdot \hat{y}$.

\begin{figure}
	\centering
	\begin{tikzpicture}[
    hit/.style={inner sep=2pt, fill, circle, black},
    simitrack/.style={shorten >= -6cm, shorten <= -6cm, opacity=.2, line width=.1cm,
        preaction={draw, line width=.3cm, opacity=.2},
        preaction={draw, line width=.5cm, opacity=.2}},
    bluesimitrack/.style={thick, blue,
        preaction={shorten >= -6cm, shorten <= -6cm, draw, opacity=.2, line width=.1cm,
            preaction={draw, blue, line width=.3cm, opacity=.2},
            preaction={draw, blue, line width=.5cm, opacity=.2}}}
        ]

    \path [use as bounding box] (-1.5,-3.8) rectangle (5,3);

    \node at (2.5,-3) [below] {$z$ axis (along the beam)};
    \node at (-1, 0) [left] {$x$};

        \begin{scope}[yshift=-3.6cm]
            \draw [latex-latex] (-1,.4) -- (-1,0) -- (5,0);
            \draw [thick, black, fill=gray]
                (-1,0) -- (-.5,0) -- (0,.4) -- (.5, 0) -- (5,0);
            \node at (0,0) [below] {\small Kernel};
        \end{scope}

    \clip (-1,-3) rectangle (5,3);

    \begin{scope}[xscale=.65, xshift=-.6cm]

        % def f(name, slope, factor):
        %     v = np.arange(1,9)
        %     vv = (v-.5)*slope + .05*np.sin(v*factor)
        %     q = np.linspace(.5, 8, 400)
        %     qq = (q-.5)*slope + .05*np.sin(q*factor)
        %     print(f'% {name} slope={slope} factor={factor}')
        %     for a,b in zip(v,vv):
        %         print(f'\coordinate ({name}{a}) at ({a}, {b:.3});')
        %     print()
        %     return q, qq, v, vv

        % (v-.5)*slope + .05*np.sin(v*factor)
        % A slope=0.4 factor=-5
        \coordinate (A7) at (1, 0.248);
        \coordinate (A6) at (2, 0.627);
        \coordinate (A5) at (3, 0.967);
        \coordinate (A4) at (4, 1.35);
        \coordinate (A3) at (5, 1.81);
        \coordinate (A2) at (6, 2.25);
        \coordinate (A1) at (7, 2.62);

        % B slope=0.06 factor=6
        \coordinate (B8) at (1, 0.016);
        \coordinate (B7) at (2, 0.0632);
        \coordinate (B6) at (3, 0.112);
        \coordinate (B5) at (4, 0.165);
        \coordinate (B4) at (5, 0.221);
        \coordinate (B3) at (6, 0.28);
        \coordinate (B2) at (7, 0.344);
        \coordinate (B1) at (8, 0.412);

        % C slope=-0.35 factor=7
        \coordinate (C7) at (1, -0.142);
        \coordinate (C6) at (2, -0.475);
        \coordinate (C5) at (3, -0.833);
        \coordinate (C4) at (4, -1.21);
        \coordinate (C3) at (5, -1.6);
        \coordinate (C2) at (6, -1.97);
        \coordinate (C1) at (7, -2.32);

    \end{scope}

    \draw[xstep=.2, ystep=10, yshift=-4cm, gray, very thin, use as bounding box] (-1,0) grid (5,8);

    \draw [simitrack] (A5) -- (A7) ;
    \draw [simitrack] (B6) -- (B8) ;
    \draw [simitrack] (C5) -- (C7) ;

    \draw [fill,white] (0,0) circle (.065) node [left] {PV};

    \node [hit] at (A1) {};
    \node [hit] at (A2) {};
    \node [hit] at (A3) {};
    \node [hit] at (A4) {};
    \node [hit] at (A5) {};
    \node [hit] at (A6) {};
    \node [hit] at (A7) {};

    \node [hit] at (B1) {};
    \node [hit] at (B2) {};
    \node [hit] at (B3) {};
    \node [hit] at (B4) {};
    \node [hit] at (B5) {};
    \node [hit] at (B6) {};
    \node [hit] at (B7) {};
    \node [hit] at (B8) {};

    \node [hit] at (C1) {};
    \node [hit] at (C2) {};
    \node [hit] at (C3) {};
    \node [hit] at (C4) {};
    \node [hit] at (C5) {};
    \node [hit] at (C6) {};
    \node [hit] at (C7) {};

\end{tikzpicture}
	\caption{The kernel for a simple three track system. 4,000 planes in $z$ (simplified in diagram) are computed to make the kernel along $z$ (bottom).}
	\label{fig:kernel}
\end{figure}
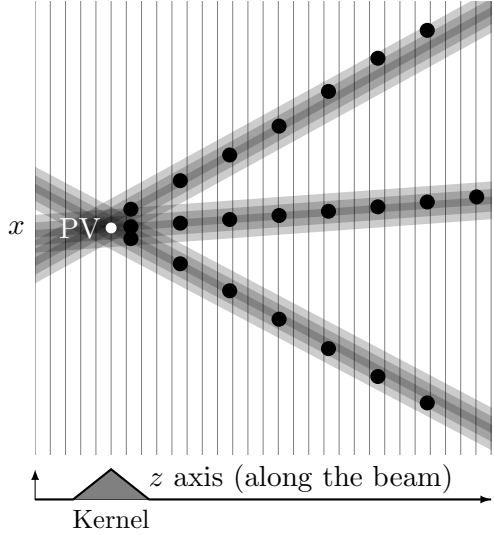

A series of 4,000 regularly spaced planes along the $z$-axis (beamline) cover the active area of the VELO in $\unit[100]{\mu m}$ increments. Each plane is divided into a course 8 by 8 grid, and the kernel value is computed in the center of each bin. Once this is done, a standard minimization algorithm is used to find the maximum point starting from the center of the bin that reported the highest value; see Figure \ref{fig:kernel}. This maximum kernel value is recorded as the value for this plane in $z$, and the procedure is repeated for all 4000 planes.

Finally, the target for the training is generated from the simulated truth information. Each PV is represented by a normalized Gaussian with a fixed width of $\unit[100]{\mu m}$. These values are recorded into 4000 bins matching the distribution of the kernel in z. This series of Gaussian pulses represents the probability of a bin containing a PV.  % TODO: Mention variable width in plans
PVs with at least 5 detectable tracks within the LHCb detector acceptance are called LHCb PVs.

\section{Network Design}

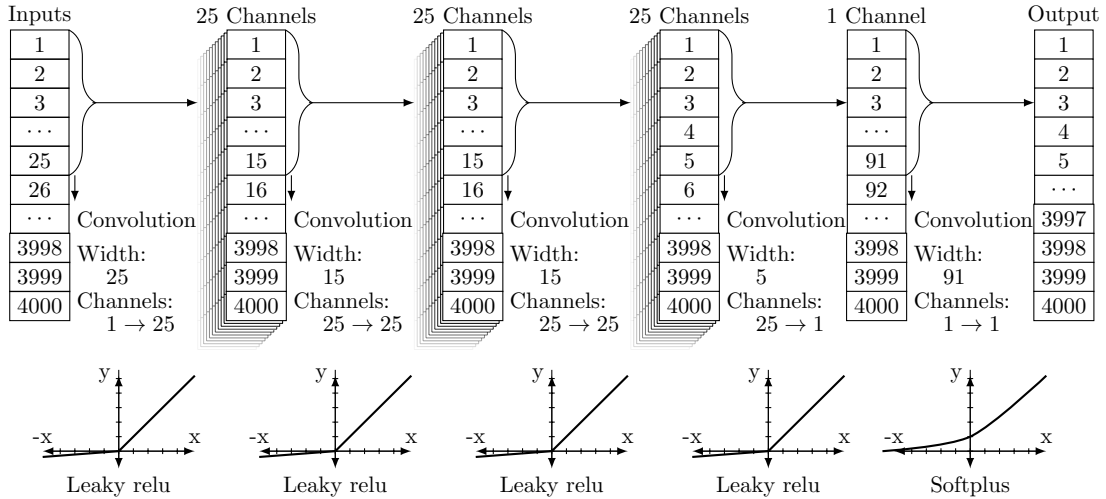
\begin{figure}
	\centering
	\newcounter{layer}
\setcounter{layer}{0}
\def\wid{3.7}

\begin{tikzpicture}[
    scale=.77,
    every node/.style={scale=.77}]

% Layer 1
\begin{scope}[xshift=\thelayer*\wid cm]
\node at (0,0) {Inputs};
\foreach \i [count=\xi] in {1,2,3,$\cdots$,25,26,$\cdots$,3998,3999,4000} {
    \node at(0, -\xi/2) [draw, minimum width = 1cm, minimum height = .5cm] {\i};
}

\draw [decorate,decoration={brace,amplitude=10pt}] (.5,-.25) -- (.5,-2.75);
\draw [-latex] (.6, -2.75) -- (.6, -3.2);
\draw [-latex] (.9, -1.5) -- (\wid - 1, -1.5);
\begin{scope}[xshift=.5cm, yshift=-3.5cm]
\node at (0, 0) [right] {Convolution};
\node at (0,  -.6) [right] {Width:};
\node at (.5, -1) [right] {25};
\node at (0, -1.4) [right] {Channels:};
\node at (.5, -1.8) [right] {$1\rightarrow25$};
\end{scope}
\end{scope}
\stepcounter{layer}

% Layer 2
\begin{scope}[xshift=\thelayer*\wid cm]
\node at (0,0) {25 Channels};
% Backing
\foreach \i in {.5,.45,.4,.35,.3,.25,.2,.15,.1,.05} {
	\begin{scope}[xshift=-.5cm-\i cm, yshift=-.25cm -\i cm]
	\path[fill=white] (0,0) rectangle (1,-5);
	\draw [opacity=1-\i*2 + .05, ystep=.5] (0,0) grid (1,-5);
	\end{scope}
}
\foreach \i [count=\xi] in {1,2,3,$\cdots$,15,16,$\cdots$,3998,3999,4000} {
	\node at(0, -\xi/2) [draw, fill=white, minimum width = 1cm, minimum height = .5cm] {\i};
}

\draw [decorate,decoration={brace,amplitude=10pt}] (.5,-.25) -- (.5,-2.75);
\draw [-latex] (.6, -2.75) -- (.6, -3.2);
\draw [-latex] (.9, -1.5) -- (\wid - 1, -1.5);
\begin{scope}[xshift=.5cm, yshift=-3.5cm]
\node at (0, 0) [right] {Convolution};
\node at (0,  -.6) [right] {Width:};
\node at (.5, -1) [right] {15};
\node at (0, -1.4) [right] {Channels:};
\node at (.5, -1.8) [right] {$25\rightarrow25$};
\end{scope}
\end{scope}
\stepcounter{layer}

% Layer 3
\begin{scope}[xshift=\thelayer*\wid cm]
\node at (0,0) {25 Channels};
% Backing
\foreach \i in {.5,.45,.4,.35,.3,.25,.2,.15,.1,.05} {
	\begin{scope}[xshift=-.5cm-\i cm, yshift=-.25cm -\i cm]
	\path[fill=white] (0,0) rectangle (1,-5);
	\draw [opacity=1-\i*2 + .05, ystep=.5] (0,0) grid (1,-5);
	\end{scope}
}
\foreach \i [count=\xi] in {1,2,3,$\cdots$,15,16,$\cdots$,3998,3999,4000} {
	\node at(0, -\xi/2) [draw, fill=white, minimum width = 1cm, minimum height = .5cm] {\i};
}

\draw [decorate,decoration={brace,amplitude=10pt}] (.5,-.25) -- (.5,-2.75);
\draw [-latex] (.6, -2.75) -- (.6, -3.2);
\draw [-latex] (.9, -1.5) -- (\wid - 1, -1.5);
\begin{scope}[xshift=.5cm, yshift=-3.5cm]
\node at (0, 0) [right] {Convolution};
\node at (0,  -.6) [right] {Width:};
\node at (.5, -1) [right] {15};
\node at (0, -1.4) [right] {Channels:};
\node at (.5, -1.8) [right] {$25\rightarrow25$};
\end{scope}
\end{scope}
\stepcounter{layer}

% Layer 4
\begin{scope}[xshift=\thelayer*\wid cm]
\node at (0,0) {25 Channels};
% Backing
\foreach \i in {.5,.45,.4,.35,.3,.25,.2,.15,.1,.05} {
	\begin{scope}[xshift=-.5cm-\i cm, yshift=-.25cm -\i cm]
	\path[fill=white] (0,0) rectangle (1,-5);
	\draw [opacity=1-\i*2 + .05, ystep=.5] (0,0) grid (1,-5);
	\end{scope}
}
\foreach \i [count=\xi] in {1,2,3,4,5,6,$\cdots$,3998,3999,4000} {
	\node at(0, -\xi/2) [draw, fill=white, minimum width = 1cm, minimum height = .5cm] {\i};
}

\draw [decorate,decoration={brace,amplitude=10pt}] (.5,-.25) -- (.5,-2.75);
\draw [-latex] (.6, -2.75) -- (.6, -3.2);
\draw [-latex] (.9, -1.5) -- (\wid - 1, -1.5);
\begin{scope}[xshift=.5cm, yshift=-3.5cm]
\node at (0, 0) [right] {Convolution};
\node at (0,  -.6) [right] {Width:};
\node at (.5, -1) [right] {5};
\node at (0, -1.4) [right] {Channels:};
\node at (.5, -1.8) [right] {$25\rightarrow1$};
\end{scope}
\end{scope}
\stepcounter{layer}

% Layer 5
\begin{scope}[xshift=\thelayer*\wid cm - .5 cm]
\node at (0,0) {1 Channel};
% Backing
%\foreach \i in {.4,.3,.2,.1} {
%\begin{scope}[xshift=-.5cm-\i cm, yshift=-.25cm -\i cm]
%\path[fill=white] (0,0) rectangle (1,-5);
%\draw [opacity=1-\i*2, ystep=.5] (0,0) grid (1,-5);
%\end{scope}
%}
\foreach \i [count=\xi] in {1,2,3,$\cdots$,91,92,$\cdots$,3998,3999,4000} {
	\node at(0, -\xi/2) [draw, fill=white, minimum width = 1cm, minimum height = .5cm] {\i};
}

\draw [decorate,decoration={brace,amplitude=10pt}] (.5,-.25) -- (.5,-2.75);
\draw [-latex] (.6, -2.75) -- (.6, -3.2);
\draw [-latex] (.9, -1.5) -- (\wid - 1, -1.5);
\begin{scope}[xshift=.5cm, yshift=-3.5cm]
\node at (0, 0) [right] {Convolution};
\node at (0,  -.6) [right] {Width:};
\node at (.5, -1) [right] {91};
\node at (0, -1.4) [right] {Channels:};
\node at (.5, -1.8) [right] {$1\rightarrow1$};
\end{scope}
\end{scope}
\stepcounter{layer}

% Layer Output
\begin{scope}[xshift=\thelayer*\wid cm - 1 cm]
\node at (0,0) {Output};
\foreach \i [count=\xi] in {1,2,3,4,5,$\cdots$,3997,3998,3999,4000} {
	\node at(0, -\xi/2) [draw, fill=white, minimum width = 1cm, minimum height = .5cm] {\i};
}
\end{scope}
\stepcounter{layer}

\begin{scope}[xshift=-.5cm, yshift=-7.5cm]

% Loop as needed
\foreach \i in {\wid * 0.5,\wid * 1.5, \wid * 2.5, \wid * 3.5} {
	\begin{scope}[xshift=\i cm]
	\draw [step=.25] (-1.2, .05) grid (1.2, -.05);
	\draw [step=.25] (-.05, -.2) grid (.05, 1.4);
	\draw [black, latex-latex] (-1.3, 0) node[above] {-x} -- (1.3, 0) node[above] {x};
	\draw [black, latex-latex] (0, -.3)  -- (0, 1.3) node[left] {y};
	\draw [thick, black](-1.3, -.1) -- (0, 0) -- (1.3,1.3);
	\node at (0, -.6) {Leaky relu};
	\end{scope}
}

\begin{scope}[xshift=\wid * 4.5 cm - .25 cm]
\draw [step=.25] (-1.2, .05) grid (1.2, -.05);
\draw [step=.25] (-.05, -.2) grid (.05, 1.4);
\draw [black, latex-latex] (-1.3, 0) node[above] {-x} -- (1.3, 0) node[above] {x};
\draw [black, latex-latex] (0, -.3)  -- (0, 1.3) node[left] {y};
\draw [thick, black, smooth]  plot coordinates {
	(-1.5, 0)
	(0, .25)
	(1.3, 1.3)};
\node at (0, -.6) {Softplus};
\end{scope}

\end{scope}

\end{tikzpicture}
	\caption{The final network architecture used for the results section. This network has four convolutional layers. For training, dropout was introduced in-between the layers, and at the beginning of the training the final layer was a Fully Connected (FC) layer, and was replaced later in the training procedure.}
	\label{fig:nnarch}
\end{figure}

The network was designed primarily using convolutional layers using PyTorch \cite{paszke2017automatic}. The most successful network to date is shown in Figure \ref{fig:nnarch}. The widths of each convolution were chosen based on visual inspection of the data; the number of channels were increased until benefits were no longer noticeable upon adding new channels. The activation function in-between each hidden layer is a leaky ReLU.

symmetric with respect to fractionally underestimating or overestimating the target values where the target values in bins with KDE = 0 are reset to $ \epsilon = 10^{-5} $ to create a mathematically tractable problem.  Ignoring the singularity when the target value is exactly zero, the initial loss function was designed to be symmetric with to

The final activation layer was initially a Sigmoid, since this can only produce values between 0 and 1, much like one would expect for a probability. However, this produced poorly shaped final probability distributions, with a distinctly square shape capped near 1, rather than Gaussian, and regularly over-estimated the probability. After changing to the Softplus function, the output probability distributions were much closer to the target. The flat derivative of the Sigmoid for large values is likely to blame for the inability of the network to correct a value that is too large.

The loss function was originally designed to be symmetric with respect to fractionally underestimating or overestimating the target values where the target values in bins with KDE = 0 are reset to $ \epsilon = 10^{-5} $ to create a mathematically tractable problem.  Ignoring the singularity when the target value is exactly zero, the initial loss function was designed to be symmetric with respect to  $r \equiv \frac{\hat y}{y}$ and $r^{-1}$, where $y$ is the predicted value and $\hat y$ is the target value. This is similar to cross entropy, which is commonly defined as
$
\mathrm{cost} = - \left(
y \ln \hat y + (1-y) \ln (1 - \hat y)
\right)
$.
Our original ``symmetric'' cost function therefore was defined as
\begin{eqnarray}
r_i & \equiv & \frac{\hat y_i + \epsilon}{y_i + \epsilon}, \\
z_i & \equiv & \frac{2 r_i}{r_i + r_i^{-1}}, \\
\mathrm{cost}  & \equiv &- \sum_\mathrm{bins} \ln z_i.
\end{eqnarray}
The parameter $ \epsilon $ is tunable; it should be $ \ll 0.01 $ to avoid confusion between bins wtih KDE = 0 and bins with finite predictions for KDE.

\begin{figure}
	\centering
	\includegraphics[width=\textwidth]{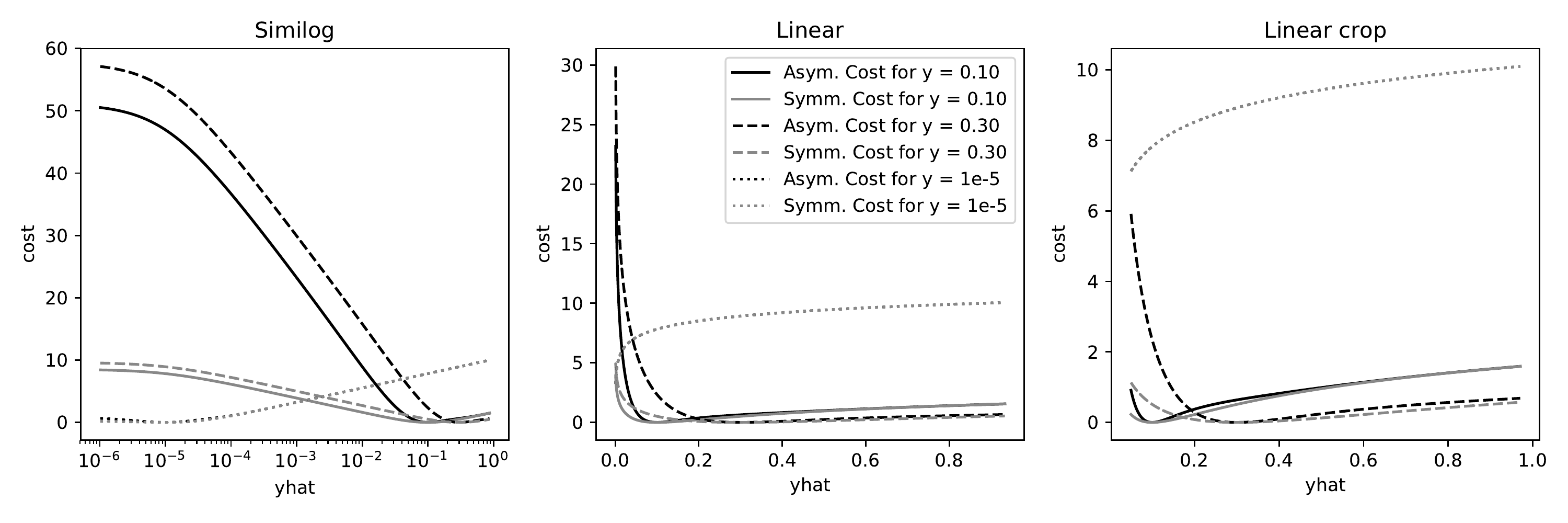}
	\caption{Plots of the loss function vs.\ $\hat y$ for several value of $y$. Log scale on the left, linear in the middle, and a reduced range linear plot on the right. Darker lines correspond to the asymmetric cost function. $\epsilon=10^{-5}$, and $a=5.0$.}
	\label{fig:loss}
\end{figure}

This loss function was found to highly favor minimal false positives (FP) at the expense of efficiency. A single asymmetry parameter was added to control the cost function, and provides a powerful control for selecting the FP to efficiency tradeoff. This loss function can be constructed from the other one with an asymmetry term $a$ and then
\begin{equation}
z_i' = z_i \left(
1+a e^{-r_i}
\right).
\end{equation}
Then $z_i'$ is used in place of $z_i$.
Figure \ref{fig:loss} shows the response for several cost functions, in both symmetric and asymmetric forms.

One of the most important additions to the original network was ``masking''. Some PVs do not pass our criterion for a good, detectable LHCb PV due to a lack of a sufficiently large number of detectable tracks. These PVs were given a masked region roughly equal to the width of the PV Gaussian probability in the target kernel, and the cost function skips these regions. This keeps the training from punishing or rewarding discovery in these regions. All further calculations, such as integrated efficiency, also ignore these masked regions.

\section{Results}

\begin{figure}
	\centering
	\includegraphics[width=\textwidth]{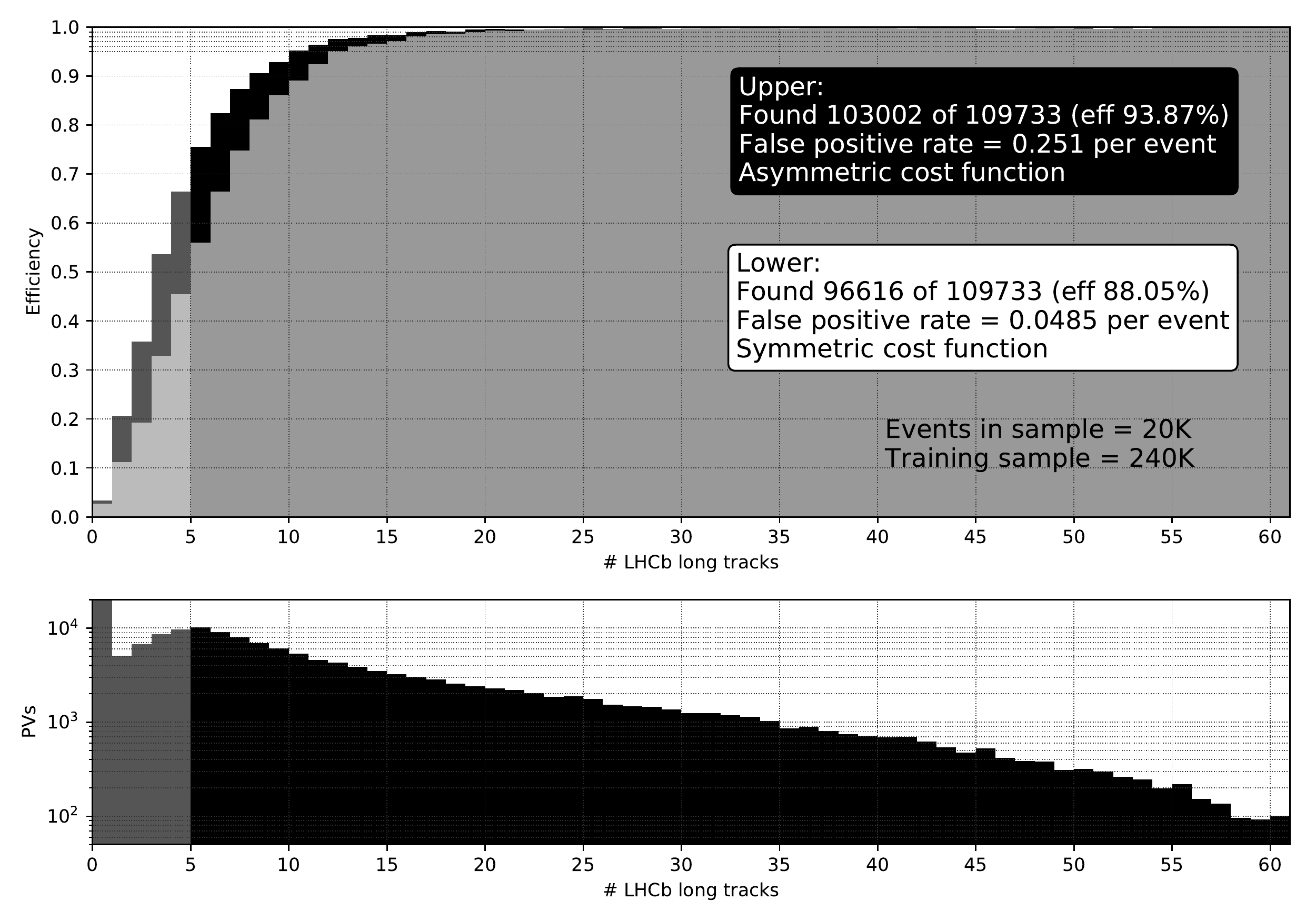}
	\caption{
		(Top plot) Results from training a network with a symmetric and an asymmetric cost function (with $a=5$).
		The lighter shaded region was trained the asymmetric cost function. The darker shaded region was trained with the asymmetric cost function.
		(Bottom plot) This histogram shows the distribution of PVs with the indicated number of tracks within the LHCb forward acceptance.
	}
    \label{fig:results}
\end{figure}

Our current results are shown in Figure \ref{fig:results}. This shows an integrated efficiency of 88\% for the symmetric cost function, and almost 94\% for a highly asymmetric cost function. Above 20 long LHCb tracks, there is nearly 100\% efficiency. False positive rates here are reported per-event.

\begin{figure}
	\centering
	\includegraphics[width=.8\textwidth]{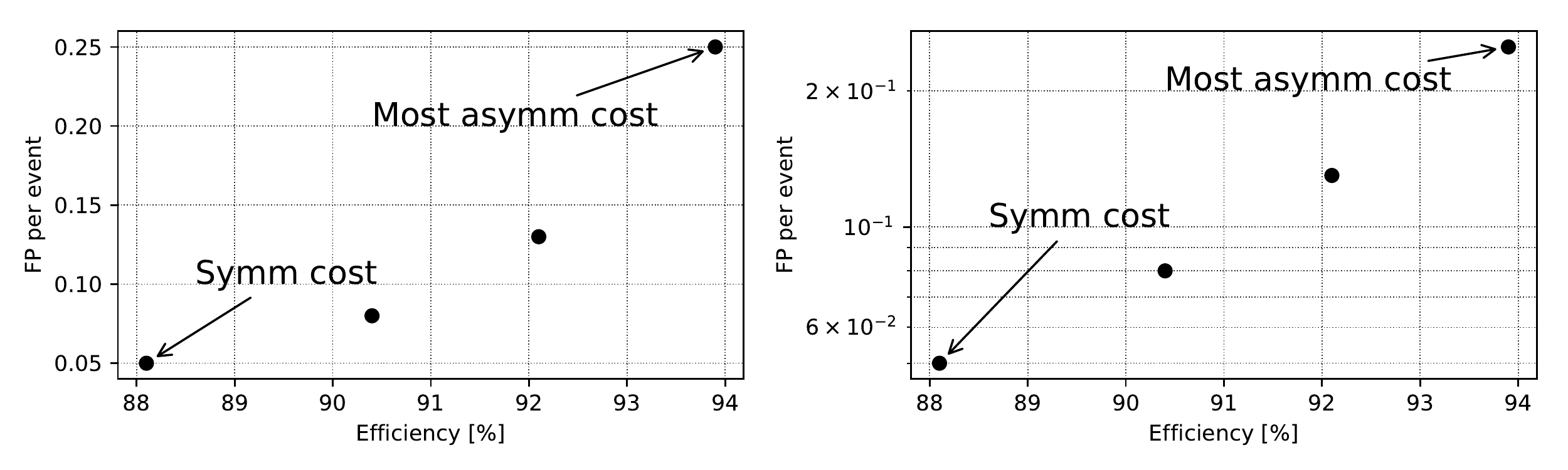}
	\caption{False positive rate vs.\ efficiency for several values of asymmetry term $a$. Log scale shown on the right, where the points are almost linear.}
	\label{fig:efffp}
\end{figure}

In Figure \ref{fig:efffp}, several different values of $a$ were used in training, up to $a=5.0$, as well as a symmetric training (equivalent to $a=0$). This shows the asymmetry parameter can be used as a tuning parameter to control this trade-off between efficiency and false positive rate.

\section{Plans}

While the initial study has been effective and has shown exciting results, there are several planned improvements. The width of each PV probability in the target distribution is currently fixed to $\unit[100]{\mu m}$; however, we could instead have a variable width based on the tracks in the PV, and study the network's ability to learn this information.

We are storing the point in $x$ and $y$ where the maximum occurred. This information can be included in the network to give the model more to learn from. The current problem with this is that the network overestimates the importance of these values compared to $z$ and stalls. This information could also be used to predict all three coordinates of the PV instead of just $z$; early studies on this have been promising. We plan to associate individual tracks with candidate PVs  probabilistically in a second step, and then combine the information to reduce the false positive rate and identify secondary tracks and secondary vertex candidates, again probabilistically.

This work has been focused so far on a 1D kernel; however, a 2D kernel is also a possibility. 1D looks to be sufficient for the upgrade conditions for LHCb, but other detectors or much higher pileup could create higher congestion in the 1D kernel which should be dramatically reduced in 2D. The kernel would be binned in one more coordinate, such as $x$ or $y$.

The next project will be to split the prototracking out from the kernel generation in our software, to allow the tracks to be input from the official LHCb simulation. There is work ongoing to do this, as well as to implement the inference engine in the LHCb software stack.

\section*{Acknowledgements}

This work was supported by the National Science Foundation under Cooperative Agreement OAC-1836650, OAC-1739772, and OAC-1740102. It was also supported by the University of Cincinnati Women in Science and Engineering program.

\section*{References}

\bibliography{pvfinder}

\end{document}